\documentclass[runningheads]{llncs}
\usepackage[dvipsnames]{xcolor}
\usepackage{hyperref}
\usepackage{subfigure}
\usepackage[ruled,vlined,linesnumbered]{algorithm2e} 
\usepackage{amsmath}
\usepackage{amssymb}
\usepackage{multirow}
\usepackage{multicol}
\usepackage{makecell}
\usepackage{graphicx}
\usepackage{courier}
\usepackage{booktabs}
\usepackage{mathrsfs}
\usepackage{cite}
\usepackage{array}
\usepackage{color}
\graphicspath{{images/}{../images/}}
\begin{document}
%

\title{Evolutionary Multi-objective Architecture Search Framework: Application to COVID-19 3D CT Classification}
\titlerunning{EMARS: Evolutionary Multi-objective Architecture Search Framework}
%


\author{
Xin He\inst{1}  \and
Guohao Ying\inst{2} \and
Jiyong Zhang$\dagger$\inst{3}\and
Xiaowen Chu$\dagger$\inst{4}\inst{1}\thanks{$\dagger$ \text{Corresponding authors (xwchu@ust.hk; jzhang@hdu.edu.cn).}}
} 

\authorrunning{Xin He, Guohao Ying, Jiyong Zhang, and Xiaowen Chu}
%
\institute{
Hong Kong Baptist University, Hong Kong, China \\
\and 
University of Southern California, CA, USA  
\and
Hangzhou Dianzi University, Hang Zhou, China 
\and The Hong Kong University of Science and Technology (Guangzhou), China
}


%
%
\maketitle              

\begin{abstract}

The COVID-19 pandemic has threatened global health. Many studies have applied deep convolutional neural networks (CNN) to recognize COVID-19 based on chest 3D computed tomography (CT). Recent works show that no model generalizes well across CT datasets from different countries, and manually designing models for specific datasets requires expertise; thus, neural architecture search (NAS) that aims to search models automatically has become an attractive solution. To reduce the search cost on large 3D CT datasets, most NAS-based works use the weight-sharing (WS) strategy to make all models share weights within a supernet; however, WS inevitably incurs search instability, leading to inaccurate model estimation. In this work, we propose an efficient \textbf{E}volutionary \textbf{M}ulti-objective \textbf{AR}chitecture \textbf{S}earch (\textbf{EMARS}) framework. We propose a new objective, namely \textbf{potential}, which can help exploit promising models to indirectly reduce the number of models involved in weights training, thus alleviating search instability. We demonstrate that under objectives of accuracy and potential, EMARS can balance exploitation and exploration, \textit{i.e.,} reducing search time and finding better models. Our searched models are small and perform better than prior works on three public COVID-19 3D CT datasets.


\end{abstract}
\section{Introduction}


The rapid spread of \textit{coronavirus disease 2019} (COVID-19) pandemic has threatened global health. Isolating infected patients is an effective way to block the transmission of the virus. Thus, fast and accurate methods to detect infected patients are crucial. Chest CT is relatively easy to perform and has been proved an important complement to nucleic acid test \cite{CT_twice}. However, there is a serious lack of radiologists during the pandemic. Many researchers have applied deep learning (DL) techniques to assist CT diagnosis. For COVID-19 3D CT classification, there are two mainstream CNN-based methods: 1) multiview-based methods \cite{iran_ctset,li2020artificial} use 2D CNN to extract features for each 2D CT slice and then fuse these features to make predictions; and 2) voxel-based methods \cite{he_aaai_covid,Zheng2020} feed 3D CNNs with 3D CT scans to make full use of the geometric information. He \textit{et al.} \cite{benchmarkcovid} benchmark a series of hand-crafted 2D and 3D CNNs and demonstrate that 3D CNNs generally outperform 2D CNNs.

Some recent works \cite{he_aaai_covid,generalization_covid} benchmark multiple COVID-19 datasets from different countries and find that no model can maintain absolute advantages on different datasets. However, since it is difficult to design models manually for specific datasets, the neural architecture search (NAS)  \cite{he2021automated,nas_survey} has become an attractive solution to discover superior models without human assistance. Reinforcement learning \cite{nas2016,enas}, gradient descent (GD) \cite{darts}, and evolutionary algorithm (EA) \cite{yang2020cars,amoebanet} are three mainstream NAS methods. The comparative results of a recent survey \cite{he2021automated} show that the EA-based NAS can discover better networks than other types of NAS methods. However, the better performance of EA-based NAS is at the cost of more computing resources because they need to retrain all searched models to compare their performance, \textit{e.g.,} AmoebaNet \cite{amoebanet} took 3,150 GPU days to search. Thanks to the weight-sharing method \cite{enas,weightsharing_survey}, any model can be evaluated without retraining, and Yang et al. \cite{yang2020cars} reduced the search time of the EA-based NAS to 0.4 GPU days. NAS was originally proposed for large-scale 2D image tasks. Although some works \cite{benchmarkcovid,he_aaai_covid} have extended NAS to search 3D models for COVID-19 3D datasets, they suffered from the search instability (analyzed in Sec. \ref{sec:potential}) incurred by weight-sharing, which leads to fluctuation in the search process and even worse results than random search in some cases. In this work, we propose an efficient \textbf{E}volutionary \textbf{M}ulti-objective  \textbf{AR}chitecture \textbf{S}earch framework, dubbed as \textbf{EMARS}. We summarize our contributions below.




\begin{enumerate}
    \item We propose a new objective, \textit{i.e.,} \textit{potential}, which can help exploit promising models and indirectly reduce the number of models involved in weights training, thereby alleviating search instability.
    \item We demonstrate that compared to conventional objective settings (\textit{e.g.,} only considering accuracy), EMARS that aims at accuracy, potential, and small size objectives can trade-off between exploitation and exploration, reducing search time by 22\% on average and discovering better models. 
    \item Our searched models are small in size and outperform prior works \cite{benchmarkcovid,he_aaai_covid} on three public datasets: CC-CCII \cite{ccccii}, MosMed \cite{mosmeddata}, and Covid-CTset \cite{iran_ctset}.
\end{enumerate}

\section{Preliminaries}

In this section, we describe the common basis of weight-sharing neural architecture search (NAS) \cite{weightsharing_survey}. NAS is formulated as a bi-level optimization problem:

\begin{equation}
\begin{array}{cl}
\min _{\alpha} & L_{\text {val }}\left(w^{*}, \alpha\right) \\
\text { s.t. } & w^{*}=\operatorname{argmin}_{w} L_{\text {train }}(w, \alpha)
\end{array}
\end{equation}

\noindent where $L_{\text {train}}$ and $L_{\text {val}}$ indicate the training and validation loss; $w$ and $\alpha$ indicate the weights and architecture of a candidate model. The early NAS methods \cite{nas2016,amoebanet} search and evaluate the networks by retraining them from scratch, resulting in huge computational cost. To reduce the burden, the weight-sharing strategy \cite{weightsharing_survey} was proposed, in which the SuperNet $\mathcal{N}$ contains all possible architectures (subnets) and its weights $\mathcal{W}$ are shared among these subnets. The architecture and weights of each subnet are denoted by $\mathcal{N}(\alpha)$ and $\mathcal{W}(\alpha)$, respectively, where $\alpha$ is the subnet architecture, encoded by one-hot sequences (described in Sec. \ref{section:search_space}). The loss of a subnet is expressed as $L(\alpha) = L(\mathcal{N(\alpha)},\mathcal{W(\alpha)},X, Y)$, where $L, X, Y$ indicate the loss function, input data, and target, respectively, and the gradient of subnet weights is $\nabla_{\mathcal{W}(\alpha)} = \frac{\partial L(\alpha)}{\partial \mathcal{W}}$. Then gradients of SuperNet weights $\mathcal{W}$ can be calculated as the average gradient of all subnets, \textit{i.e.,}  $\nabla_{\mathcal{W}} = \frac{1}{N}\sum_{i=1}^N\nabla_{\mathcal{W}(\alpha_i)} = \frac{1}{N}\sum_{i=1}^N\frac{\partial L(\alpha_i)}{\partial \mathcal{W}}$,  where $N$ is the total number of subnets. Obviously, it is not practical to use all subnets to update SuperNet weights at each time. Therefore, we use a mini-batch of subnets for training, detailed as Eq. \ref{eq:approx}




\begin{equation}
    \nabla_{\mathcal{W}}\approx \frac{1}{M}\sum_{i=1}^M\nabla_{\mathcal{W}(\alpha_i)}
    \label{eq:approx}
\end{equation}

\noindent where $M$ is the number of subnets sampled in a mini-batch and $M<<N$. In our experiments, we find that $M=1$ works just fine, \textit{i.e.,} we can update $\mathcal{W}$ using the gradient from any single sampled subnets for each training batch.

\section{Methodology}




\subsection{Potential objective: Alleviating Search Instability} \label{sec:potential}


%
By \textit{instability}, we mean that the same subnet can produce a completely different performance at different times of the search process. The instability is caused by the weight-sharing  strategy because the weights of all subnets are coupled, then an update of any subnet's weights is bound to affect (usually negatively) other subnets. Therefore, the performance of a subnet at a specific time does not necessarily represent its real performance but instead misleads the direction of evolutionary search (described in Sec. \ref{sec:seach_alg}). To mitigate the search instability caused by weight-sharing, a natural idea is to reduce the number of models involved in weights training (\textit{i.e.,} Eq. \ref{eq:approx}). For this reason, some works \cite{anglenas,oneshot_shrink} directly reduce the number of models by progressively shrinking the search space based on the model performance, but this may eliminate promising models in the early stage of the search. To avoid this problem, we take an indirect approach in which we keep exploring various models in the early stage of the search and then spend more effort on training those promising models in the later stage of the search. In this way, we can indirectly reduce the number of models involved in weights training without deliberately reducing the search space. However, how do we determine whether a model is promising or not?

Here, we propose a new objective, namely \textit{potential}, to help find promising models. Specifically, for each sampled model, we maintain and update its historical performances $Z=(E,F)$, where $E=[e_1,...,e_m]^T$ is a column vector recording the epochs when the model is sampled, $F=[f_1,...,f_m]^T$ is a column vector recording the corresponding validation accuracy. Note that, $Z$ is dynamically updated with the search process, so the size of $Z$ (\textit{i.e.,} $m$) varies for models. The potential $\mathcal{P}$ of a model is calculated by ordinary least squares (OLS):

\begin{equation}
\mathcal{P} = (E^TE)^{-1}E^TF
\end{equation}


\noindent To some extent, $E$ can also reflect how promising a model is, \textit{e.g.,} if $E$ is densely distributed, it means this model outperforms other models in multiple rounds of search and hence wins more chances to be sampled. However, considering only $E$ will exacerbate the Matthew effect, and the search may get trapped in a local optimum. Our proposed potential solves this problem by considering the coupling relation between sampling frequency $E$ and validation accuracy $F$, \textit{i.e.,} the growing trend of accuracy rather than the accuracy at a specific time. The larger the $\mathcal{P}$ value, the more promising the model is. 




 

\subsection{Evolutionary Search}\label{sec:seach_alg}





\begin{figure}[!ht]
    \centering
    \includegraphics[width=\textwidth]{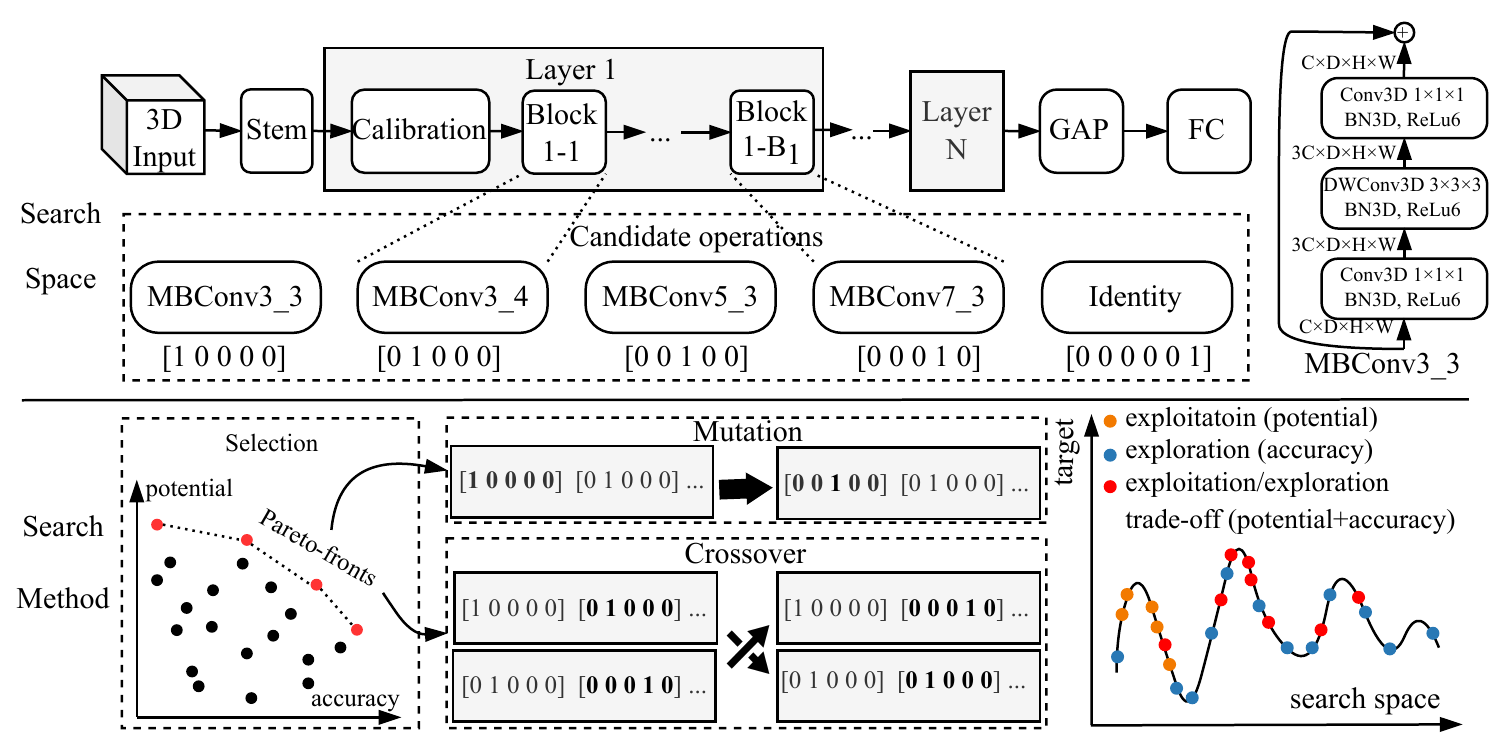}
    \caption{
    Overview of search space and search method. Upper-right: MBConv$3\_3$, where C, D, H, W indicate channels, depth, height, and width. Lower-right: An example of exploitation and exploration under different objectives. (best viewed in color)
    }
    \label{fig:search_space}
\end{figure}

The search algorithm (see Supplement Alg. 1) starts with a warm-up stage, followed by the evolutionary search stage. In the warm-up, the SuperNet is trained by uniformly sampling subnets, thus all candidate operations are trained equally. After the warm-up, top-$P$ best-performing subnets form the initial population, \textit{i.e.,} $\mathcal{A}^{(0)}$, and will be evolved for multiple generations. Each generation comprises two sequential processes: 1) \textit{weights training}, where each individual (\textit{i.e.,} subnet) is selected from the population and trained based on Eq \ref{eq:approx}; and 2) \textit{architecture search}, comprising selection, crossover, and mutation (see Fig. \ref{fig:search_space}).

\textbf{Selection.} After weights training, we record multiple objectives for all individuals in the population. We adopt NSGA-II \cite{nsga2} method to select Pareto-front individuals under the recorded objectives from the population. We compare different combinations of these objectives in Sec. \ref{sec:results} and find that searching with potential and accuracy can discover better models with less cost.

\textbf{Crossover\&Mutation.} The selection produces $K$ Pareto-front individuals, based on which we further generate $P-K$ new individuals. Each new individual is generated by randomly sampling from the SuperNet or performing crossover and mutation (CM) with certain probabilities. The basic unit of CM is the one-hot sequence, representing the candidate operation (see Fig. \ref{fig:search_space}). 



\textbf{Exploitation\&Exploration.} Fig. \ref{fig:search_space} (lower-right) shows an example of two important issues in the evolutionary algorithm (EA) based search: \textit{exploration} and \textit{exploitation}. Exploitation prefers the current optimal solution, which reduces search cost but may lead to a local optimum; exploration is more likely to find the optimal solution but consumes more resources. The common opinion about EA is that the steps of crossover and mutation determine the exploration, and exploitation is done by selection. However, our experiments in Sec. \ref{sec:results} show that setting different objectives in the selection step can also control the evolution direction.  Specifically, accuracy and potential will make the evolution process towards exploration and exploitation, respectively, while combining accuracy and potential can balance exploration and exploitation.


\subsection{Search Space}\label{section:search_space}

\textbf{SuperNet.} The search space is represented by a SuperNet $\mathcal{N}$, containing all possible subnets. SuperNet comprises two parts: 1) the searchable part, \textit{i.e.,} $N=6$ layers; 2) the fixed part, \textit{i.e.,} stem block, global average pooling \cite{GAP}, and a fully connected layer. The stem block is a standard $3\times3\times3$ 3D convolution followed by a 3D batch normalization and a ReLu6 activation function \cite{relu6}.

\textbf{Layer.} The $i$-th layer comprises a calibration block and $B_i$ searchable blocks. The calibration block is a 3D $1\times1\times1$ point-wise convolution to solve the problem of feature dimension mismatch; thus, all subsequent blocks have a stride of 1. The number of searchable blocks and the stride of calibration block in six layers are [4,4,4,4,4,1] and [2,2,2,1,2,1], respectively. The output channels of the stem block and six layers are 32 and [24,40,80,96,192,320], respectively.

\textbf{Block.} Each searchable block is a candidate operation, encoded by a one-hot sequence. We adopt eight candidate operations, including a \textit{skip-connection} operation and seven mobile inverted bottleneck convolutions \cite{mobilenetv2}, denoted by \textit{MBConv}$k\_e$, where $k\_e\in\{3\_3,3\_4,3\_6,5\_3,5\_4,7\_3,7\_4\}$, $k$ is the kernel size of the intermediate depth-wise convolution (DWConv), and $e$ is the expansion ratio between the input channel and inner channel of MBConv. 


\section{Experiments}\label{sec:exp_imp}


\subsection{Implementation Details}

\textbf{Datasets.} For a fair comparison, we apply the same three datasets as prior works \cite{benchmarkcovid,he_aaai_covid}. CC-CCII \cite{ccccii} has 3,993 CT scans of three classes: novel coronavirus pneumonia (NCP), common pneumonia (CP), and normal case; MosMed \cite{mosmeddata} has 1,110 scans of NCP and normal classes; Covid-CTset \cite{iran_ctset} has 526 scans of NCP and normal classes. More details of datasets can be referred to supplement.



\textbf{Search stage.} We use four Nvidia V100 GPUs to search for 100 epochs, where the warm-up stage has 10 epochs. During each search epoch, a population of models are equally trained on the training set and evaluated on the validation set. The population size is 20, where 10 Pareto-front models are selected from the population using NSGA-II \cite{nsga2} under multiple objectives (\textit{e.g.,} validation accuracy, potential, and model size), and 10 new models are generated by crossover and mutation with the probabilities of 0.3 and 0.2. To improve search efficiency, we set the input size ($width\times height\times depth$) to $64\times64\times16$. We use Adam optimizer \cite{adam} with a weight decay of 3e-4 and an initial learning rate of 0.001.

\textbf{Retraining stage.} After the search stage, we combine the training and validation set and retrain the Pareto-front models on the combined set for 200 epochs. We use the same Adam settings as the search stage. The 3D input sizes of CC-CCII, MosMed, and Covid-CTset datasets are $128\times128\times32$, $256\times256\times40$, and $512\times512\times32$, respectively. Our framework is based on NNI \cite{nni} and available at: \url{https://github.com/marsggbo/MICCAI2022-EMARS}.


\subsection{Results and Analysis}\label{sec:results}

\begin{figure}[!ht]
    \centering
    \includegraphics[width=\textwidth]{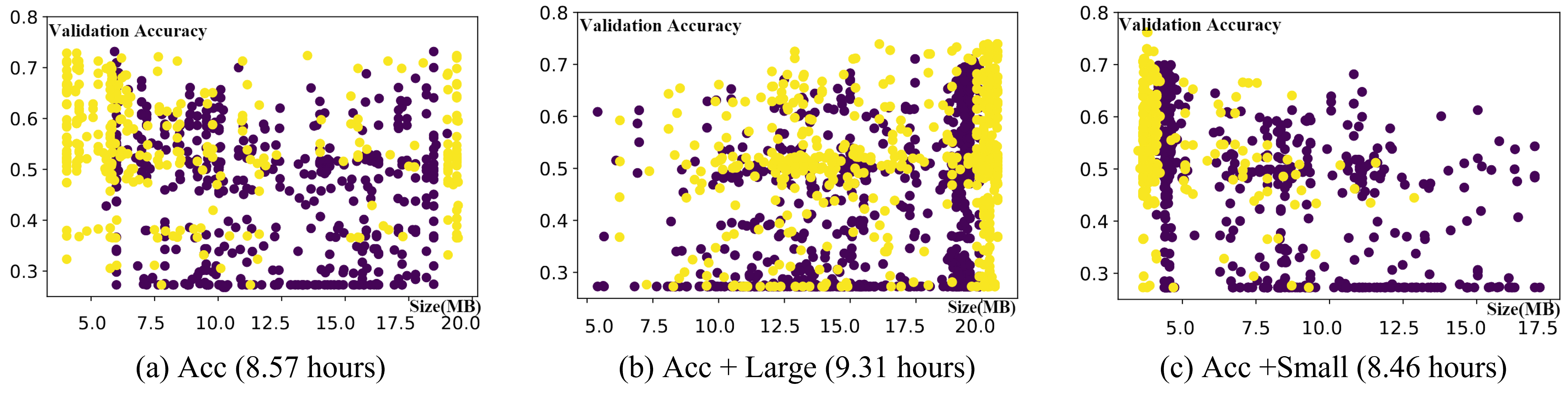}
    
    \caption{
    The model size-aware search results. X and Y axes indicate model size and validation accuracy (Acc). The purple and yellow points indicate the sampled models in the first and last half of the search stage, respectively. (best viewed in color)
    }
    \label{fig:model_size}
\end{figure}


\begin{figure}[!ht]
    \centering
    \includegraphics[width=\textwidth]{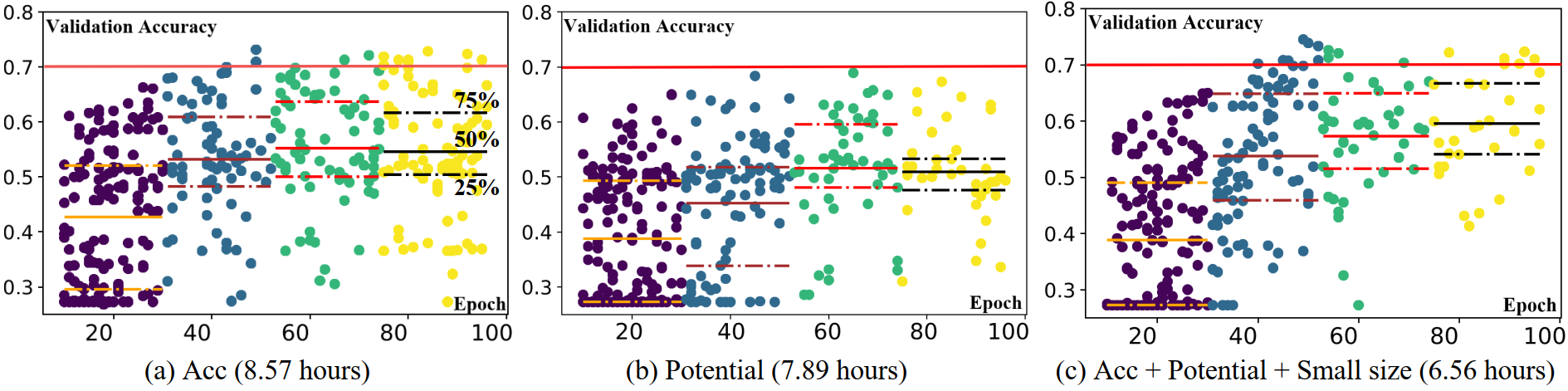}
    \caption{
    The potential (P) aware search results. Different colored points indicate the models sampled in different epoch periods. The solid and dashed lines in each period indicate the average and 25/75 percentile accuracy, respectively. (best viewed in color)
    }
    \label{fig:potential}
\end{figure}

\textbf{Model Size-aware Search.} Fig. \ref{fig:model_size} presents model size-aware search results on CC-CCII dataset. Fig. \ref{fig:model_size} (a) shows that searching under only validation accuracy (Acc) will explore both extremes of model size, but with no performance gain, while Fig. \ref{fig:model_size} (b)\&(c) show that additional consideration of model size on top of Acc helps find better models in the later stage, indicating multi-objective can facilitate the search process. Besides, compared to Fig. \ref{fig:model_size} (b), searching under Acc and small model size in Fig. \ref{fig:model_size} (c) can not only reduce search time from 9.31 hours to 8.46 hours but also discover competitive models.

\textbf{Potential-aware Search.} We further build three experiments on the CC-CCII dataset to validate \textit{potential} objective. Each sub-figure of Fig. \ref{fig:potential} divides the search process into four periods based on the search epoch. Each period is presented with different colors and marked with the accuracy of 25/50/75th percentiles. Fig. \ref{fig:potential} (a) shows that searching under Acc tends to \textit{explore} more models, regardless of whether the model performance is good or bad, leading to wasting time on those unpromising models (lower-right points). On the contrary, in Fig. \ref{fig:potential} (b), the difference between the 25th and 75th percentiles and the number of sampled models are gradually reduced with the search process, which implies that potential will guide the evolution process in the later stage to \textit{exploit} promising models already discovered. Although it reduces search time, it has lower Acc due to being trapped in local optima in the early stage. Fig. \ref{fig:potential} (c) shows that searching under potential, Acc, and small size can reduce the search time by 19\% on average and balance exploitation and exploration. Specifically, the first two periods are dominated by exploration, as a wide accuracy range of models is explored, and we can find models with an accuracy of more than 0.7 faster in the second period. On the other hand, the last two periods focus more on exploitation, as the number of unpromising models is significantly reduced, and the accuracy of 25/50/75th percentiles is improved steadily.




\textbf{Comparison with Prior Works.}  Table. \ref{table:base_vs_emars} compares our searched models with prior works based on four widely used metrics: accuracy, precision, sensitivity, and f1 score. Precision and sensitivity are a pair of negatively correlated metrics, so they cannot fully describe model performance. F1 score is the harmonic mean of the precision and sensitivity; thus, it is a better metric. As can be seen, our models searched under APS (accuracy, potential, and small model size) objectives have small sizes and outperform all prior hand-crafted and NAS-based models on three datasets in terms of accuracy, precision, and f1 score. Besides, MosMed is an imbalanced dataset, and we can find that the models (\textit{e.g.,} CovidNet3D-S/L and EMARS-A) searched without potential are overfitted on positive class (\textit{i.e.,} NCP), as they have extremely high sensitivity but low precision. On the contrary, EMARS-P and EMARS-APS are searched with potential objective, balancing precision and sensitivity well and achieving higher accuracy and f1 scores. More results can be referred to the Supplement.

\begin{table*}[!ht]
    \centering
    \caption{Results on CC-CCII \cite{ccccii}, MosMed \cite{mosmeddata}, and Covid-CTset \cite{iran_ctset} datasets. A, P, and S in our model name indicate accuracy, potential, and small model size, \textit{e.g.,} EMARS-A indicates the model searched under the accuracy objective.}
    \scalebox{1}[1]{
    \begin{tabular}{c|ccccccc}%
    \hline
    Dataset& Model& Size (MB) &  Type & \makecell{Accuracy} & \makecell{Precision} & \makecell{Sensitivity} & \makecell{F1} \\ \hline
    \multirow{10}{*}{\makecell{CC-\\CCII\\{[}China{]}\\\cite{ccccii}}} & ResNet3D101 \cite{videoresnet} & 325.21 &  \multirow{5}{*}{Manual}   & 85.54 & 89.62 & 77.15 & 82.92 \\  
     & DenseNet3D121 \cite{densenet3d} & 43.06 &     & 87.02 & 88.97 & 82.78 & 85.76 \\  
     & MC3\_18 \cite{videoresnet} & 43.84 &    & 86.16 & 87.11 & 82.78 & 84.89 \\ 
     &  COVID-AL \cite{covid_al}  &-  & & 86.60 & - & - & - \\
     & VGG16-Ensemble\cite{covid_ensemble} & -  & & 88.12 & 84.04 & 89.19 &86.54 \\
    \cline{2-8}
     & CovidNet3D-S \cite{he_aaai_covid} & 11.48&  \multirow{6}{*}{Auto}   & 88.55 & 88.78 & \textbf{91.72} & 90.23 \\
     & CovidNet3D-L  \cite{he_aaai_covid}& 53.26 &    & 88.69 & 90.48 & 88.08 & 89.26 \\
     & MNas3DNet \cite{benchmarkcovid} & 22.91 & & 87.14 & 88.44 & 86.09 & 87.25 \\ 
     & \textbf{EMARS-A} & 5.93 &   & \textbf{89.67} &	89.26 & 89.22 & 89.23 \\
     & \textbf{EMARS-P} & 5.63 &   & 88.78&	88.81&	88.22&	88.51 \\
     & \textbf{EMARS-APS} & 3.38 &    & 89.61 & \textbf{91.48} & 89.97 & \textbf{90.72} \\\hline\hline

    \multirow{10}{*}{\makecell{Mos-\\Med\\{[}Russia{]}\\\cite{mosmeddata}}} & ResNet3D101 \cite{videoresnet} & 325.21 & \multirow{5}{*}{Manual} & 81.82 & 81.31 & 97.25 & 88.57 \\  
     & DenseNet3D121 \cite{densenet3d} & 43.06  & & 79.55 & 84.23 & 92.16 & 88.01 \\  
     & MC3\_18 \cite{videoresnet} & 43.84   & & 80.4 & 79.43 & 98.43 & 87.92 \\ 
     & DeCoVNet \cite{decovnet} & -   & & 82.43 & - & - &- \\ 
     \cline{2-8}
     & CovidNet3D-S \cite{he_aaai_covid} & 12.48 & \multirow{5}{*}{Auto}  & 81.17 & 78.82 & 99.22 & 87.85 \\
     & CovidNet3D-L \cite{he_aaai_covid} & 60.39 &  & 82.29 & 79.50 & 98.82 & 88.11 \\
    & \textbf{EMARS-A} & 2.89 &    & 80.98 & 77.91 & \textbf{99.61} & 87.44 \\
    & \textbf{EMARS-P} & 18.22 & & 84.34 & \textbf{93.56} & 85.49 & 89.34 \\
     & \textbf{EMARS-APS} &  10.69 &  & \textbf{88.09} & 93.52 & 90.59 & \textbf{92.03} \\\hline\hline

    \multirow{10}{*}{\makecell{Covid-\\CTset\\{[}Iran{]}\\\cite{iran_ctset}}} & ResNet3D101 \cite{videoresnet} & 325.21  & \multirow{5}{*}{Manual} & 93.87 & 92.34 & 95.54 & 93.92 \\  
     & DenseNet3D121 \cite{densenet3d} & 43.06 &   & 91.91 & 92.57 & 92.57 & 92.57  \\  
     & MC3\_18 \cite{videoresnet} & 43.84 &  & 92.57 & 90.95 & 94.55 & 92.72  \\ 
     & CovCTx\cite{chetoui2021efficient_covid} & - &  & 96.37 & - & 97.00 & - \\
     & Vit-32$\times$32 \cite{vit_covid} & - & & 95.36 & - & 83.00 & - \\
     \cline{2-8}
     & CovidNet3D-S \cite{he_aaai_covid} & 8.36 & \multirow{6}{*}{Auto}   & 94.27 & 92.68 & 90.48 & 91.57 \\
     & CovidNet3D-L \cite{he_aaai_covid} & 62.82  & & 96.88 & 97.50 & 92.86 & 95.12 \\
     & \makecell{AutoGluon model \cite{autogluon_covid19}} & 93.00 & & 89.00 & 90.00 & 88.00 & 88.00 \\    
    & \textbf{EMARS-A} & 8.36 &   & 95.16 & 95.77 & 95.16 & 95.46  \\
    & \textbf{EMARS-P} & 14.41 & & 92.87 & 92.73 & 92.74 & 92.74 \\
     & \textbf{EMARS-APS} & 9.95 &  & \textbf{97.66} & \textbf{97.61} & \textbf{97.58} & \textbf{97.59} \\\hline

    \end{tabular}}
    \label{table:base_vs_emars}
\end{table*}

\section{Conclusion and Future Work}\label{sec:conclusion}

In this work, we introduce an EA-based neural architecture search (EMARS) framework, which can efficiently discover superior 3D models under multiple objectives for COVID-19 3D CT classification. We demonstrate that our proposed objective, \textit{i.e.,} \textit{potential}, can effectively alleviate the search instability and help exploit promising models. The models searched by EMARS  under accuracy and potential objectives have small sizes and outperform the previous work on three public datasets. We believe our framework can also be extended to other types of datasets and tasks (\textit{e.g.,} segmentation), which is also our future work.

\section{Acknowledge}

This work was supported in part by Hong Kong Research Matching Grant RMGS2019\_1\_23, the Zhejiang Province Nature Science Foundation of China under Grant LZ22F020003, and the HDU-CECDATA Joint Research Center of Big Data Technologies under Grant KYH063120009. 
%
%
\bibliographystyle{splncs04}
\bibliography{paper960}

\end{document}